\documentclass[aps,preprint,showpacs]{revtex4}%
\usepackage{amsfonts}
\usepackage{amsmath}
\usepackage{amssymb}
\usepackage{graphicx}%
\begin{document}
\title{Quantum transport in a resonant tunnel junction coupled to a nanomechanical oscillator}
\author{M.~Tahir}
\altaffiliation[Permanent address: ]{Department of Physics, University of Sargodha, Sargodha, 40100, Pakistan}

\email{m.tahir06@imperial.ac.uk, m.tahir@uos.edu.pk}
\author{A.~MacKinnon}
\affiliation{Department of Physics, The Blackett Laboratory, Imperial College London, South
Kensington Campus, London SW7 2AZ, U.K.}
\thanks{M.Tahir would like to acknowledge the support of the Pakistan Higher Education
Commission (HEC).}
\date{\today}


\begin{abstract}
We discuss the quantum transport of electrons through a resonant tunnel
junction coupled to a nanomechanical oscillator at zero temperature. By using
the Green's function technique we calculate the transport properties of
electrons through a single dot strongly coupled to a single oscillator. We consider a finite chemical potential difference between the right and
left leads. In addition to the main resonant peak of electrons on the dot, we
find satellite peaks due to the creation of phonons. These
satellite peaks become sharper and more significant with increasing coupling strength between the electrons and the oscillator. We also consider the energy
transferred from the electrons to the oscillator.

\end{abstract}
\maketitle

\section{Introduction}
In recent years, there has been great interest in quantum transport through both
single\cite{1} electronic levels in quantum dots and single molecules\cite{2}.
Very interesting and significant signatures of the electron phonon interaction have been observed experimentally\cite{3} in cases where the electron--phonon coupling is beyond the validity of simple perturbation theory. Single
molecular electronics\cite{4} has also attracted much attention, motivated both by the scientific challenges and by their potential applications in nanoelectromechanical systems (NEMS)\cite{5,6}. The most
interesting challenges are to verify the uncertainty principle\cite{7}, to
study the quantum transport properties as atomic scale quantum
effects\cite{8,9} become more important and significant, and the fabrication
of devices on a nano scale that are expected to be faster, more
reliable and more sophisticated than existing technology. An interesting
aspect of NEMS is the interplay between electrons, phonons and the coupling of
leads to the system\cite{9,10}. Inelastic tunneling spectroscopy\cite{11,12}
is the most direct way of observing the inelastic effects in the
current--voltage (I-V) characteristics of NEMS. NEMS can be used as
ultra-sensitive detectors for mass\cite{13}, extremely weak forces\cite{14},
ultra small displacements\cite{15} and even in chemical\cite{16} and
biological\cite{17} applications. At very low bias voltage resonances occur
with the frequency of the nanomechanical oscillator. The characteristic
frequency associated with these systems is of the order of $1\,\mbox{GHz}$%
\cite{18}. Such a high resonance frequency is sufficient to enable the
cooling\cite{19} of a nanomechanical resonator to its ground state: a
necessary condition for these measurements and something on which experimental
effort is now under way. Moreover, the quantum transport requires very highly
sensitive measurements such as are achievable using single electron
transistors (SET) or superconducting SETs (SSET).

Most of the theoretical work on transport in NEMS has been done within the
scattering theory approach (Landauer) but it disregards the contacts and their
effects on the scattering channel as well as effect of electrons and phonons
on each other\cite{20}. Very recently, the non--equilibrium Green's function
(NEGF) approach\cite{21} has been growing in importance in the quantum
transport of nanomechanical systems\cite{22}. An advantage of this method is
that it treats the infinitely extended reservoirs in an exact way, which may
lead to a better understanding of the essential features of NEMS. The
pioneering work on NEGF was done by Datta and Lake\cite{23}. NEGF has been
applied in the study of shot noise in chain models\cite{24} and disordered
junctions\cite{25} while noise in Coulomb blockaded Josephson junctions has been
discussed within a phase correlation theory approach\cite{26}. The case of an
inelastic resonant tunneling structure, in which strong electron-phonon
coupling is often considered, a very strong source-drain voltage is expected
for which  coherent electron transport in molecular devices has been
considered by some workers\cite{27} within the scattering theory approach.
Inelastic effects on the transport properties have been studied in connection
with NEMS and substantial work on this issue has been done, again within the
scattering theory approach\cite{20}. Phonon assisted tunneling of
non--resonant systems has mostly been shown by experiments on inelastic
tunneling spectroscopy (ITS). With the advancement of modern technology, as
compared to ITS, scanning tunneling spectroscopy (STS) and scanning tunneling
microscopy (STM) have proved more valuable tools for the investigation and
characterization of molecular systems\cite{28} in the conduction regime.
Recently, phonon assisted resonant tunneling conductance has been discussed
within the NEGF technique at zero temperature\cite{29}.

In the present work, we employ the Green's function method in order to discuss
the transport properties of NEMS. This is a fully quantum mechanical
formulation whose basic approximations are very transparent, as the technique
has already been used to study transport properties in a wide range of
systems. The main differences between existing work and ours are: in most of
the existing literature a very large chemical potential difference is considered
while we are able to include a range from very small to very large. In our calculation the
inclusion of the oscillator is not perturbative as the STS
experiments\cite{9,10} are beyond the range of perturbation theory. Hence, an approach is required beyond the quantum master
equation\cite{22,30} or linear response. In addition, we aim in future to apply the same
methodology to describe more clearly non--perturbative systems such as a
quantum shuttle\cite{22}. Hence, our work provides an exact analytical
solution to the current--voltage, conductance, coupling of leads with the
system, very small chemical potential difference and includes both the right
and left Fermi level response regimes. However, our theory does rely on the
commonly used wide--band approximation\cite{31,32,33} where the coupling between
leads and dot is taken to be independent of energy. This provides a way to
perform transient transport calculations from first principles while retaining
the essential physics of the electronic structure of the dot and the leads.
Another advantage of this method is that it treats the infinitely extended
reservoirs in an exact way in the present system, which may give a better
understanding of the essential features of NEMS in a more appropriate quantum
mechanical picture.

\section{Formulation}
We consider a single quantum dot connected to two identical metallic leads.
A single oscillator is coupled to the electrons on the dot and an applied
gate voltage is used to tune the single level of the dot. In the present
system, we neglect the spin degree of freedom and electron-electron
interaction effects and consider the simplest possible model system. In this initial work we also
neglect the effects of finite electron temperature of the lead reservoirs and
damping of the oscillator. Our model consists of the individual entities such
a single quantum dot and the left and right leads in their ground states at
zero temperature. The Hamiltonian of our simple system\cite{29,31,32} is%
\begin{equation}
H_{\mbox{\scriptsize dot-ph}}=\left[  \epsilon_{0}+\alpha(b^{\dagger
}+b)\right]  c_{0}^{\dag}c_{0}+\omega_{0}b^{\dagger}b\,,\label{eq:1}%
\end{equation}
where $\epsilon_{0}$ is the single energy level of electrons on the dot with
$c_{0}^{\dag},c_{0}$ the corresponding creation and annihilation operators,
$\alpha$ is the coupling strength between electrons on the dot and an
oscillator of frequency $\omega_{0}$ and $b^{\dagger},b$ are the raising and
lowering operators of the phonons. The remaining elements of the Hamiltonian
are
\begin{align}
H_{\mbox{\scriptsize leads}}  & =\sum_{k=L,R}\epsilon_{k}c_{k}^{\dagger}%
c_{k}\label{eq:2a}\\
H_{\mbox{\scriptsize leads-dot}}  & =\frac{1}{\sqrt{N}}\sum_{k=L,R}%
V_{k}\left(  c_{k}^{\dagger}c_{0}+c_{0}^{\dagger}c_{k}\right)
\,,\label{eq:2b}%
\end{align}
where $N$ is the total number of states in each lead. The total Hamiltonian of
the system is thus%
\begin{equation}
H=H_{\mbox{\scriptsize dot-ph}}+H_{\mbox{\scriptsize leads}}%
+H_{\mbox{\scriptsize leads-dot}}\,.\label{eq:3}%
\end{equation}
We write the eigenfunctions of $H_{\mbox{\scriptsize dot-ph}}$ in a $k$--space representation as%
\begin{align}
\Phi_{n}(k,x_{0}\neq0)  & =A_{n}\exp[-{\textstyle\frac{1}{2}}k^{2}%
]H_{n}(k)\exp[-{\mathrm{i}}kx_{0}]\label{eq:4a}\\
\Phi_{n}(k,x_{0}=0)  & =A_{n}\exp[-{\textstyle\frac{1}{2}}k^{2}]H_{n}%
(k)\,,\label{eq:4b}%
\end{align}
for the occupied, $x_{0}\not =0$ and unoccupied, $x_{0}=0$, dot respectively,
where $x_{0}$ is the displacement of the oscillator due to the coupling to the
electron and $H_{n}(k)$ are the usual Hermite polynomials. Here we have used
the fact that the harmonic oscillator eigenfunctions have the same form in
both real and Fourier space.

In order to transform between the representations for the occupied and
unoccupied dot we require the matrix with elements
\begin{equation}
A_{mn} = \int\Phi_{n}^{\ast}(k,x_{0}\neq0)\Phi_{n}(k,x_{0}=0)\,{\mathrm{d}}k
\label{eq:5}%
\end{equation}
which may be simplified\cite{34} as%
\begin{align}
A_{mn}  & = \frac{l}{\sqrt{\pi2^{m+n}n!m!}}\int\exp\left( -k^{2}\right)
H_{m}(k)H_{n}(k)\exp\left( \mathrm{i}kx_{0}\right) \,{\mathrm{d}}k\nonumber\\
& = \sqrt{\frac{2^{n-m} m!}{n!}} \exp\left( -{\textstyle\frac{1}{4}}x_{0}%
^{2}\right)  \left( {\textstyle\frac{1}{2}}\mathrm{i} x_{0}\right) ^{n-m}
L_{m}^{n-m}\left( {\textstyle\frac{1}{2}}{x_{0}^{2}}\right)  \label{eq:6}%
\end{align}
for $m\leq n$, where $x_{0}=\Delta/{\omega_{0}}$, $\Delta=\alpha^{2}%
/{\omega_{0}}$ and $L_{m}^{n}(x)$ are the associated Laguerre polynomials.
Note that the integrand is symmetric in $m$ and $n$ but the integral is only
valid for $m\leq n$. Clearly the result for $m>n$ is obtained by exchanging
$m$ and $n$ in equation~(\ref{eq:6}) to obtain
\begin{equation}
A_{mn} =\sqrt{\frac{2^{|n-m|}\min[m,n]!}{\max[m,n]!}}\exp\left(
-\textstyle\frac{1}{4}x_{0}^{2}\right) \left( \textstyle\frac{1}{2}%
\mathrm{i}x_{0}\right) ^{|n-m|}L_{\min[m,n]}^{|n-m|}\left( \textstyle\frac
{1}{2}{x_{0}^{2}}\right) \,. \label{eq:7}%
\end{equation}

\section{Green's functions and quantum transport}

In order to calculate analytical results and to discuss the numerical spectrum
of the quantum transport properties of nanomechanical systems, our focus in
this section is to derive an analytical relation for the current, and the
differential conductance, as a function of applied bias. In obtaining these
results we use the wide--band approximation where the self--energy of the dot
due to each lead is considered to be energy independent and is given by%
\begin{eqnarray}
\Sigma_{K}^{r} & = & n_{\mathrm{D}} \left| V_{K}\right| ^{2}\int_{-\infty
}^{+\infty} \frac{\mathrm{d}\epsilon_{K}}{E-\epsilon_{K}}\nonumber\\
& = &-2\pi\mathrm{i}n_{\mathrm D} \left| V_{K}\right| ^{2}\\
& = &-\mathrm{i}%
{\mathchoice{{\textstyle{\frac12}}}{{\textstyle{\frac12}}}{{\scriptstyle{1/2}}}{{\scriptscriptstyle{1/2}}}}%
\Gamma_{K}\,, \nonumber\label{eq:8}%
\end{eqnarray}
where $n_{\mathrm D}$ is the constant number density of the leads, $K=L,R$
represent the left and right leads and $\Gamma_{K}$ is the damping factor
($\Gamma_{L}=\Gamma_{R}=\Gamma$). Similarly
\begin{equation}
\Sigma_{K}^{a} = \left[ \Sigma_{K}^{r}\right] ^{*} = +\mathrm{i}%
{\mathchoice{{\textstyle{\frac12}}}{{\textstyle{\frac12}}}{{\scriptstyle{1/2}}}{{\scriptscriptstyle{1/2}}}}%
\Gamma_{K}\,.\label{eq:9}%
\end{equation}

We solve Dyson's equation using $H_{\mbox{\scriptsize dot-leads}}$ as a
perturbation. For the more general systems we aim to treat in future this is a reasonable small parameter.  In the present case, however, we can find an exact solution. The retarded and advanced
Green's functions on the dot, with the phonon states in the representation of
the unoccupied dot, may be written as
\begin{equation}
G_{nn^{\prime}}^{\mathrm r(a)}= \sum_{m} A_{nm}g_{m}^{\mathrm r(a)}%
A_{n^{\prime}m}^{\ast}\,,\label{eq:11}%
\end{equation}
where $g_{n}^{\mathrm{r(a)}}$ is the retarded (advanced) Green's function on
the occupied dot,%
\begin{equation}
g_{n}^{\mathrm r(a)} =\left[ E-\epsilon_{0}%
-(n+{\mathchoice{{\textstyle{\frac12}}}{{\textstyle{\frac12}}}{{\scriptstyle{1/2}}}{{\scriptscriptstyle{1/2}}}}%
)\omega_{0}+\Delta\pm\mathrm{i}\Gamma\right] ^{-1} \label{eq:12}%
\end{equation}
with $\Delta=\alpha^{2}/{\omega_{0}}$.

The lesser self--energy may be written as%
\begin{equation}
\Sigma_{K}^{<}(E)=\mathrm{i}\Gamma f_{K}(E)\,. \label{eq:13}%
\end{equation}
where  $f_{L(R)}$ are the Fermi distribution functions of the
left and right leads, which have different chemical potentials under a voltage bias, including a contribution from the state of the oscillator.

For the present case of zero temperature the lesser self--energy may be recast
in terms of the Heaviside step function $\theta(x)$ as%
\begin{equation}
\Sigma_{K}^{<}(E)=\mathrm{i}\Gamma\theta\left(  \epsilon_{\mathrm{F}%
K}%
+{\mathchoice{{\textstyle{\frac12}}}{{\textstyle{\frac12}}}{{\scriptstyle{1/2}}}{{\scriptscriptstyle{1/2}}}}%
\omega_{0}-E\right)\delta_{n,0}  \,, \label{eq:14}%
\end{equation}
where $\epsilon_{\mathrm{F}K}$ is the Fermi energy on lead $K$, and the Kronecker delta, $\delta_{n,0}$, signifies that the oscillator is initially in its ground state, $n=0$. Similarly one
can calculate the greater self--energy as
\begin{equation}
\Sigma_{K}^{>}(E)=-\mathrm{i}\Gamma\lbrack1-f_{K}(E)]\,. \label{eq:14a}%
\end{equation}
The lesser Green's function is related to the density matrix through
\begin{equation}
\rho_{nn^{\prime}}=-2\mathrm{i}G_{nn^{\prime}}^{<}\,, \label{eq:15}%
\end{equation}
Here $G_{nn^{\prime}}^{<}$ is the full lesser Green's function including the
dot and the leads. With the help of the density matrix formulation, the
current formula is%

\begin{equation}
I=<\hat{I}>=\mathop{\mbox{Tr}}\nolimits\left(  \rho\hat{I}\right)
=\mathrm{i}%
{\mathchoice{{\textstyle{\frac12}}}{{\textstyle{\frac12}}}{{\scriptstyle{1/2}}}{{\scriptscriptstyle{1/2}}}}%
\mathop{\mbox{Tr}}\nolimits\left(  G^{<}\hat{I}\right)\,,\label{eq:15a}
\end{equation}
where $\hat{I}$ is the current operator. Using this formula, we calculate the
current from the contact through both barriers and the central region with the
oscillator on the dot.
The general expression for the current is%

\begin{equation}
I=\frac{e}{4\pi}\int\left\{  \mathop{\mbox{Tr}}\nolimits\left[  \left(
\Sigma_{L}^{<}-\Sigma_{R}^{<}\right)  \left(  G^{r}-G^{a}\right)  \right]
+\mathop{\mbox{Tr}}\nolimits\left[  \left(  \left(  \Sigma_{L}^{a}-\Sigma
_{L}^{r}\right)  -\left(  \Sigma_{R}^{a}-\Sigma_{R}^{r}\right)  \right)
G^{<}\right]  \right\}  \,\mathrm{d}E\,. \label{eq:15b}
\end{equation}
We can obtain the same result by calculating the current from the dot into one
of the leads, which gives%

\begin{equation}
I=\frac{e}{4\pi}\int\{\mathop{\mbox{Tr}}\nolimits[-\Sigma_{R}^{<}\left(
G^{r}-G^{a}\right)  ]-\mathop{\mbox{Tr}}\nolimits[\left(  \Sigma_{R}%
^{a}-\Sigma_{R}^{r}\right)  G^{<}]\}\,\mathrm{d}E\,, \label{eq:16}%
\end{equation}
where the first term in the above expression vanishes for $n>0.$ The lesser
Green's function in the presence of the oscillator is given by%
\begin{equation}
G^{<}=G^{r}\Sigma^{<}G^{a}\qquad\mbox{with}\quad\Sigma^{<}=\Sigma_{L}%
^{<}+\Sigma_{R}^{<}\,. \label{eq:17}%
\end{equation}

Using the same damping factor in each lead ($\Gamma_{L}=\Gamma_{R}=\Gamma$),
the final expression for the total current ($I$) reduces to%

\begin{equation}
I=\frac{e}{4\pi}\int\mathop{\mbox{Tr}}\nolimits\left(  \Sigma_{L}^{<}%
-\Sigma_{R}^{<}\right)  \left(  G^{r}-G^{a}\right)  \,\mathrm{d}E\,. \label{eq:18}%
\end{equation}

\section{Average Energy}

To calculate the energy transferred from the electrons to the oscillator we
return to equation (\ref{eq:16})~and note that the contributions to the trace with $n>0$
may be identified with the current due to those electrons which have been
scattered inelastically with the creation of $n$ phonons. As the lesser
self--energy factors in the 1st term are zero for $n>0,$ the inelastic
contributions are solely contained in the 2nd term. The first term in equation
(\ref{eq:16}) does contribute to the total current calculated in equation (\ref{eq:18}) but does
not contribute to the energy of the oscillator. We may therefore use the
lesser Green's function (or density matrix) to calculate the energy
transferred to the oscillator to obtain
\begin{equation}
E_{\mbox{\scriptsize ph}}=\sum_{n}n\omega_{0}Y_{n}\biggl/I\qquad\text{where
}Y_{n}%
={\mathchoice{{\textstyle{\frac12}}}{{\textstyle{\frac12}}}{{\scriptstyle{1/2}}}{{\scriptscriptstyle{1/2}}}}%
\mathrm{i}\frac{\Gamma e}{4\pi}\int G_{nn}^{<}\,\mathrm{d}E\,,\label{eq:19}%
\end{equation}
From equation~(\ref{eq:19}) we may write the lesser Green's function in terms of the
lesser self--energy and the retarded and advanced Green's functions as%
\begin{equation}
G_{nn}^{<}=G_{n0}^{r}(\Sigma_{0,L}^{<}+\Sigma_{0,R}^{<})G_{0n}^{a}\,,\label{eq:20}%
\end{equation}
where we note that, as we are working at $T=0$ the self--energy terms are only
non--zero for the zero phonon state. Hence we have
\begin{align}
Y_{n} &
={\mathchoice{{\textstyle{\frac12}}}{{\textstyle{\frac12}}}{{\scriptstyle{1/2}}}{{\scriptscriptstyle{1/2}}}}%
\mathrm{i}\frac{\Gamma e}{4\pi}\sum_{m,k}\int_{-\infty}^{\infty}\left[
\frac{A_{nm}A_{0m}^{\ast}}{E-\epsilon_{0}-\left(
m+{\mathchoice{{\textstyle{\frac12}}}{{\textstyle{\frac12}}}{{\scriptstyle{1/2}}}{{\scriptscriptstyle{1/2}}}}%
\right)  \omega_{0}+\Delta+\mathrm{i}\Gamma}\right]  \nonumber\\
&  \qquad\qquad\times\biggl[\mathrm{i}\Gamma\theta\left(  \epsilon
_{\mathrm{F}L}%
+{\mathchoice{{\textstyle{\frac12}}}{{\textstyle{\frac12}}}{{\scriptstyle{1/2}}}{{\scriptscriptstyle{1/2}}}}%
\omega_{0}-E\right)  +\mathrm{i}\Gamma\theta\left(  \epsilon_{\mathrm{F}%
R}%
+{\mathchoice{{\textstyle{\frac12}}}{{\textstyle{\frac12}}}{{\scriptstyle{1/2}}}{{\scriptscriptstyle{1/2}}}}%
\omega_{0}-E\right)  \biggr]\nonumber\\
&  \qquad\qquad\times\left[  \frac{A_{0k}A_{nk}^{\ast}}{E-\epsilon_{0}-\left(
k+{\mathchoice{{\textstyle{\frac12}}}{{\textstyle{\frac12}}}{{\scriptstyle{1/2}}}{{\scriptscriptstyle{1/2}}}}%
\right)  \omega_{0}+\Delta-\mathrm{i}\Gamma}\right]  \,\mathrm{d}%
E\,.\label{eq:21}%
\end{align}
We note, however, that this expression is non--zero even when $\epsilon
_{\mathrm{F}L}=\epsilon_{\mathrm{F}R}$ and $\epsilon_{0}<\epsilon_{\mathrm{F}%
}$ due to the dot being permanently occupied in these circumstances. To remove
this term we subtract the contribution when the 2 Fermi levels are equal. This
reduces the expression for $Y_{n}$ to
\begin{equation}
Y_{n}%
=-{\mathchoice{{\textstyle{\frac12}}}{{\textstyle{\frac12}}}{{\scriptstyle{1/2}}}{{\scriptscriptstyle{1/2}}}}%
\frac{\Gamma^{2}e}{4\pi}\int_{\epsilon_{\mathrm{F}R}}^{\epsilon_{\mathrm{F}L}%
}\left\vert \sum_{m}\frac{A_{nm}A_{0m}^{\ast}}{E-\epsilon_{0}-\left(
m+{\mathchoice{{\textstyle{\frac12}}}{{\textstyle{\frac12}}}{{\scriptstyle{1/2}}}{{\scriptscriptstyle{1/2}}}}%
\right)  \omega_{0}+\Delta+\mathrm{i}\Gamma}\right\vert ^{2}\,\mathrm{d}%
E\,.\label{eq:22}%
\end{equation}
After integrating the above expression \cite{35}, we arrive at the final result%
\begin{align}
Y_{n} &
=-{\mathchoice{{\textstyle{\frac12}}}{{\textstyle{\frac12}}}{{\scriptstyle{1/2}}}{{\scriptscriptstyle{1/2}}}}%
\frac{\Gamma^{2}e}{4\pi}\sum_{m,k}\frac{A_{n,m}A_{0,m}^{\ast}A_{0,k}%
A_{n,k}^{\ast}}{(k-m)\omega_{0}+2\mathrm{i}\Gamma}\nonumber\\
&  \times\left\{  \ln\left[  \frac{\epsilon_{\mathrm{F}L}-\epsilon_{0}%
-m\omega_{0}+\Delta-\mathrm{i}\Gamma}{\epsilon_{\mathrm{F}L}-\epsilon
_{0}-k\omega_{0}+\Delta+\mathrm{i}\Gamma}\right]  -\ln\left[  \frac
{\epsilon_{\mathrm{F}R}-\epsilon_{0}-m\omega_{0}+\Delta-\mathrm{i}\Gamma
}{\epsilon_{\mathrm{F}R}-\epsilon_{0}-k\omega_{0}+\Delta+\mathrm{i}\Gamma
}\right]  \right\}  \,.\label{eq:23}%
\end{align}
Hence, the average energy transferred to the oscillator may be calculated
using equation (\ref{eq:19}).

\section{Discussion of Results}
\begin{figure}[htb]
\begin{center}
\includegraphics[width=\textwidth]{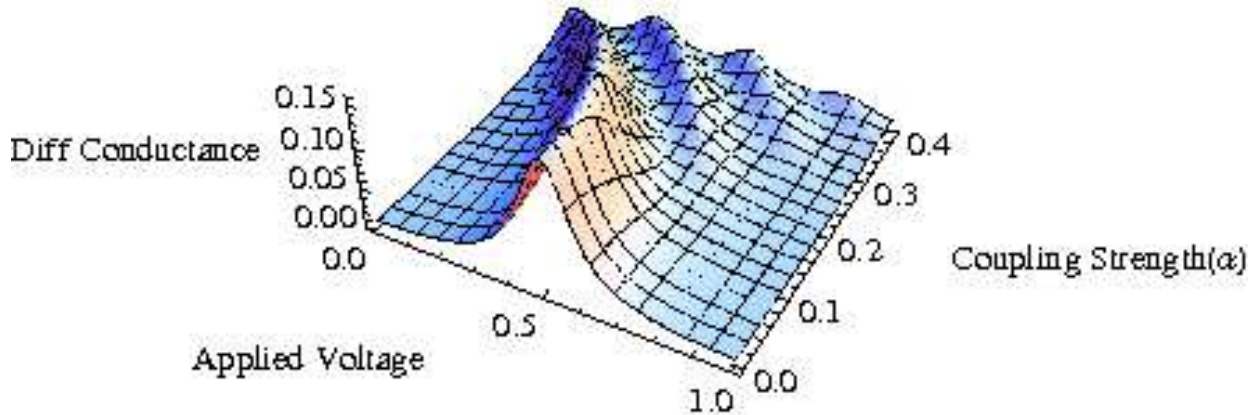}
\end{center}
\caption{Differential conductance (dimensionless) as a function of applied
voltage $\epsilon_{\mathrm{F}L}$ (in arbitrary units) and coupling strength
$\alpha$. Gate voltage $\epsilon_{0}=0.5$, oscillator frequency $\omega
_{0}=0.3$, self--energy $\Gamma= 0.3\omega_{0}$. }%
\label{figure1}%
\end{figure}
The differential conductance is shown graphically in figure~\ref{figure1} as a function of applied
voltage for different values of coupling strength, using the same parameters
as \cite{29,31,32,38,39,40}: the single energy level of the dot  $\epsilon_{0}=0.5$,
the characteristic frequency of the oscillator  $\omega_{0}=0.3$, the damping
factor $\Gamma=0.3\omega_{0}$ and the chemical potentials $0 \le
\epsilon_{\mathrm{F}L} \le1$ and $\epsilon_{\mathrm{F}R}=0$. These are chosen to illustrate the physics of such systems rather than to represent a specific implementation. The oscillator
induced resonance effects are clearly visible in the
numerical results. It must be noted that we have obtained these results in the
regime of strong and zero or weak coupling of the oscillator with the
electrons on the dot. The coupling between the leads and the dot is considered
to be symmetric and we assume that the leads have constant density of states. With increasing coupling strength, the number of satellite peaks
also increases while for zero or weak coupling we find only the basic resonance. This confirms the effect of the coupling between the
electrons on the dot and the single oscillator mode where higher energy electrons are able to drop to the dot energy by creation of phonons. We note the similarity of figure~\ref{figure1} to figure~3a of \cite{41}, which refers, however, to the transmission amplitude of an interference device, albeit using a similar Hamiltonian.  Transport processes involving creation or annihilation of phonons are a common feature of NEMS.

Closer analytical examination of the expression for the differential conductance (\ref{eq:22}) shows that the main resonance
peaks occur when the applied voltage, $\epsilon_{FL}$ is equal to the energy eigenvalues of the
coupled dot electron and oscillator. The main peak $(n=0)$ is
given by the Lorentzian form with its center at the 
$\epsilon_{FL}=\epsilon_{0}-\Delta$, 
known as a Breit-Wigner\cite{36,37,38} resonance.
The satellite peaks due to the emission of phonons can be seen
on the positive energy side with $\epsilon_{FL}=\epsilon_{0}-\Delta+n\omega_{0}$ where $\omega_{0}$ is the characteristic frequency
of the oscillator.

The main or basic resonance peak is the elastic or zero phonon transition. The
amplitude of the satellite peaks or steps is much smaller than the basic
resonance peak. The electrons that tunnel onto the dot can only excite the
oscillator mode as at zero temperature there are no phonons available to
be absorbed. Moreover, we have seen that with increasing coupling strength,
the number and intensity of the satellite peaks increases but their intensity
always remains much smaller than the main peak. The peaks or steps
in the current characteristics vanish if the upper electrochemical potential is
smaller than the dot energy plus the oscillator frequency. 

\begin{figure}[tbh]
\begin{center}
\includegraphics[width=0.6\textwidth]{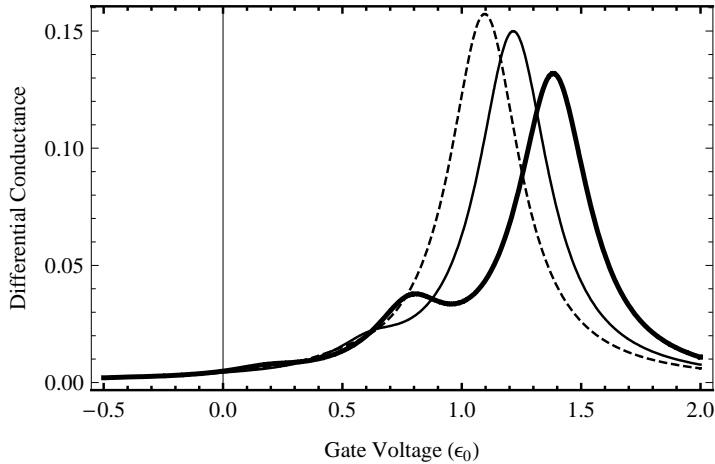}
\end{center}
\caption{Differential conductance (dimensionless) as a function of gate
voltage $\epsilon_{0}$, with applied voltage $\epsilon_{\mathrm{F}L} = 1$,
oscillator frequency $\omega_{0}=0.6$, self energy $\Gamma=0.3\omega_{0}$ and
coupling strength $\alpha=0.4\omega_{0} (\mbox{dotted line}), 0.6\omega
_{0}(\mbox{light solid line}), 0.8\omega_{0} (\mbox{bold solid line})$. }%
\label{figure2}%
\end{figure}
The differential conductance as a function
of gate voltage, $\epsilon_0$, is shown in fig.~\ref{figure2} for various coupling strengths at $T=0$.  The main peak at
$\epsilon_{0}=\epsilon_{\mathrm{F}L}+\Delta)$ corresponds to elastic or
zero--phonon transition and the satellites peaks are due to emission of
phonons corresponding to $n=1,2,3,4,\ldots$. This shows more and more
satellites corresponding to every multiple of $\omega_{0}$.

With increasing coupling strength while keeping the temperature zero, we see
that the energy transferred to the oscillator increases with increasing
coupling strength, while the amplitude of the satellite peaks is much smaller
than the main peak which is shifted toward the right by a factor $\Delta$. The
amplitude of the main peak is also affected: its magnitude decreases with increasing coupling strength.

\begin{figure}[tbh]
\begin{center}
\includegraphics[width=0.45\textwidth]{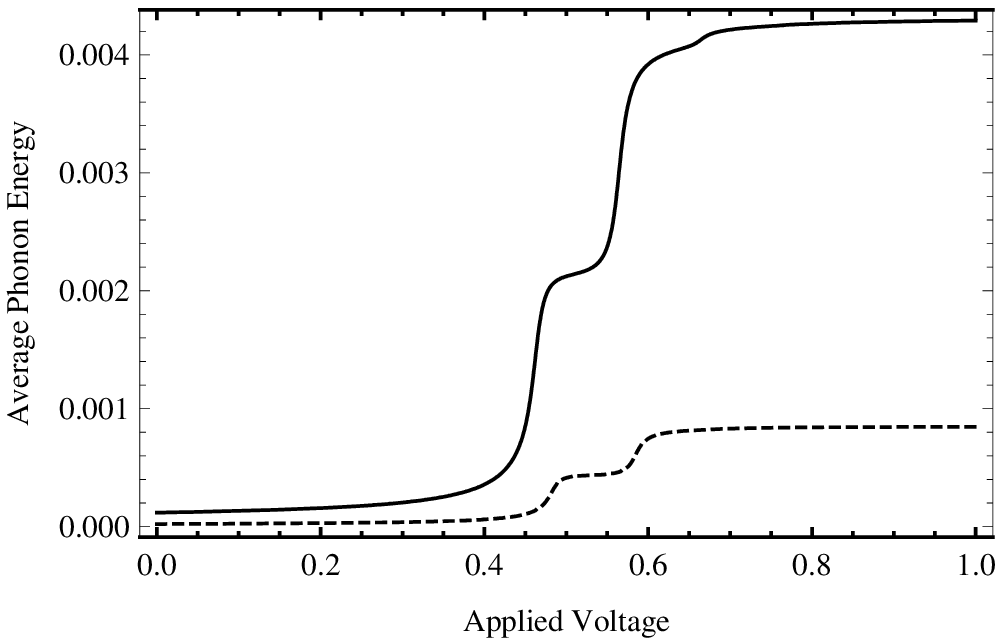}\hfill
\includegraphics[width=0.45\textwidth]{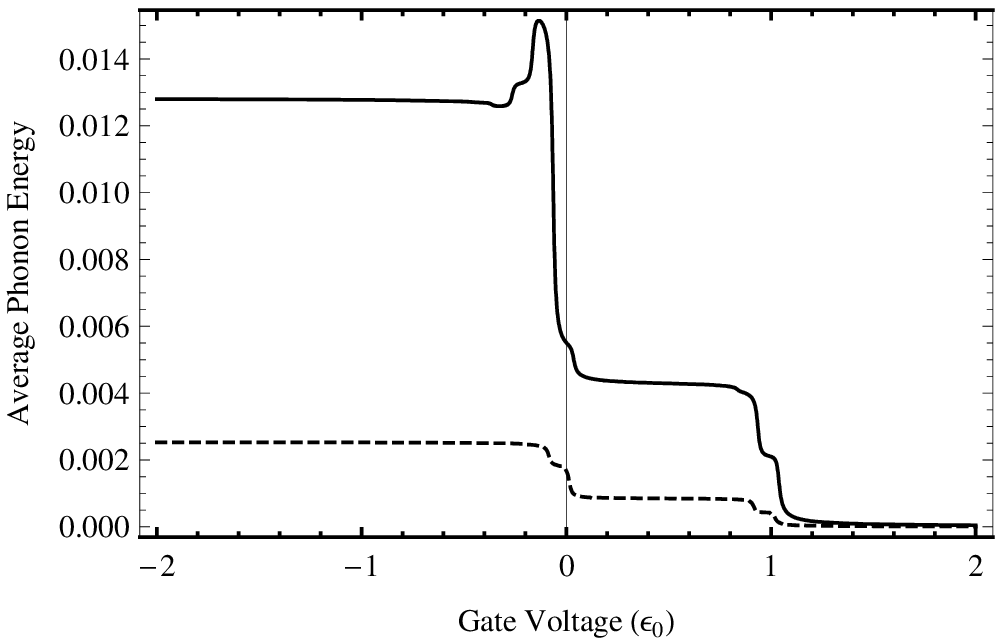}
\end{center}
\caption{ Average Energy transferred to the oscillator as a function of applied voltage $\epsilon_{\mathrm{F}L}$ with gate voltage $\epsilon_0 = 0.5$ (left figure) and as a function of gate voltage $\epsilon_0$ with $\epsilon_{\mathrm{F}L} = 1$ (right figure). The oscillator frequency $\omega_0 = 0.1$, self–energy $\Gamma= 0.1\omega_0$, and coupling strength $\alpha = 0.4\omega_0$ (dotted line) and $\alpha = 0.6\omega_0$ (solid line). }%
\label{figure34}%
\end{figure}
In fig.~\ref{figure34}, we plot the average energy that is being
transferred to the oscillator per transmitted electron as a function of applied bias
and gate voltage. At zero temperature the oscillator can only gain energy from
the electrons. We note that there is some structure as individual phonons are
excited but there is also a saturation level $\propto\alpha^{4}$. The peak
just below $\epsilon_{0}\approx\epsilon_{\mathrm{F}R}$ in the right hand figure is
due to the fact that there is no elastic transmission in this regime and
\textit{all\/} transmitted electrons result in the creation of phonons. Note
that this is the average energy transferred when the system starts in its
ground state and should not be confused with the energy transferred after many
electrons have interacted with the oscillator. We consider $T=0$ in this work for simplicity and will consider finite temperature effects later.
Moreover, the phonon energy
of the oscillator on the dot and the level width are both typically larger
than the experimental temperature. Our theory is in good agreement with the
growing body of theoretical\cite{29,31,32,38} and experimental\cite{39,40}
work in this area.

\section{Summary}
In this work, we analyzed the dynamics of a nanomechanical oscillator coupled
to a resonant tunnel junction by using the Green's function approach without
treating the electron phonon coupling as a perturbation. We have derived an
expression for the current and differential conductance and discuss it in
detail for different values of the coupling strength. We have found
steps/peaks in the current spectrum as a function of the chemical potential
difference in addition to the main resonant step, due to the transfer of
energy from electrons on the dot to the oscillator. We have also studied the
effect of gate voltage. We also derive an expression for the average energy
transferred from the electrons to the oscillator. We have shown that the steps
grow with increasing coupling strength of electrons on the dot and the
oscillator. This confirms that the additional satellite peaks or steps
in the spectrum of numerical results are due to the transfer of energy from
the electrons to the oscillator.

\end{document}